\documentclass[12pt,preprint]{aastex}

\begin{document}

\title{
The Inclination Angle of and Mass of the Black Hole in XTE~J1118+480
}

\author{Dawn M. Gelino \email{dawn@ipac.caltech.edu}}
\affil{Michelson Science Center, Caltech, MS 100-22, 770 South Wilson Avenue, Pasadena, CA, 91125}
\author{\c{S}\"olen Balman, \"Umit K{\i}z{\i}lo\u{g}lu, Arda Y{\i}lmaz 
}
\affil{Middle East Technical University, Ankara, Turkey}
\author{Emrah Kalemci}
\affil{Sabanc\i\ University, Orhanl\i-Tuzla, Istanbul, 34956, Turkey}
\author{John A. Tomsick}
\affil{Center for Astrophysics and Space Sciences, Code 0424, University of California at San Diego, La Jolla, CA, 92093}

\begin{abstract}
We have obtained optical and infrared photometry of the quiescent soft X-ray transient XTE J1118+480. In addition to optical and $J$-band variations, we present the first observed $H$- and $K_s$-band ellipsoidal variations for this system. We model the variations in all bands simultaneously with the WD98 light curve modeling code. The infrared colors of the secondary star in this system are consistent with a K7V, while there is evidence for light from the accretion disk in the optical. Combining the models with the observed spectral energy distribution of the system, the most likely value for the orbital inclination angle is $68^{\circ}\pm2^{\circ}$.  This inclination angle corresponds to a primary black hole mass of $8.53\pm0.60 \,M_{\odot}$. Based on the derived physical parameters and infrared colors of the system, we determine a distance of 1.72$\pm$0.10 kpc to XTE J1118+480.  
\end{abstract}


\section{Introduction}

Transient low mass X-ray binaries (LMXBs) exhibit large and abrupt
X-ray and optical outbursts that can be separated by decades of
quiescence \citep{csl97}.  For these systems, the compact object is
a black hole or a neutron star, and the companion is normally a low-mass
K- or M-type dwarf-like star \citep{cc03}.  During their periods of
quiescence, these systems are faint at X-ray, optical, and infrared
(IR) wavelengths.  While the quiescent X-ray emission can be caused
by accretion onto a compact object or thermal emission from the
surface of a neutron star \citep{garcia01,wijnands04}, the companion
can dominate the luminosity at optical and IR energies.  During the
binary orbit, the changing aspect of the tidally-distorted companion
causes a periodic modulation of the optical and IR emission
\citep{gelino01}. Measurements of these ``ellipsoidal variations''
provide information about the physical parameters of the binary.

XTE J1118+480 ($\alpha_{2000}$ = 11$^{\rm h}$18$^{\rm m}$10.85$^{\rm s}$, 
$\delta_{2000}$ = 48$^{\circ}$02$\arcmin$12.9$\arcsec$) was discovered with the all-sky monitor (ASM) on the {\it Rossi X-Ray Timing Explorer} by 
\citet{rem00} on 2000 March 29, while \citet{gar00} spectroscopically identified its 12.9 magnitude optical counterpart. Owing to its position in the Galactic halo, this high latitude ({\it b} = +62$^{\circ}$) system has been observed by numerous groups over many wavelength regimes.  Recent orbital parameters determined from optical spectra suggest an orbital period of 4.078 hr and a secondary star radial velocity semi-amplitude of 709$\pm$7 km s$^{-1}$ \citep{tor04}. These values imply a mass function of {\it f(M)}=6.3$\pm$0.2 M$_{\odot}$, identifying the compact object as a black hole. 

Determining a precise black hole mass requires an accurate measurement of the orbital 
inclination angle of the system.  As discussed in 
 \citet{gho01}, the 
best way to find the inclination angle in a non-eclipsing system is to model 
its infrared ellipsoidal light curves. In the IR regime, there is a smaller chance of contamination from other sources of light in the system.  While modeling several light curves from one wavelength regime helps to constrain model parameters, simultaneously modeling light curves that span more than one wavelength regime provides tighter constraints than modeling IR light curves alone.  Previous inclination estimates for XTE J1118+480 have come from modeling optical ellipsoidal variations as the system approached quiescence \citep{mcc01, wag01, zur02}.  These inclination angles range from 55$^{\circ}$\ \citep{mcc01} to 83$^{\circ}$\ \citep{wag01}, and correspond to primary masses of 10 M$_{\odot}$ and 6.0 M$_{\odot}$, respectively. Since this system has been known to exhibit optical superhumps from the precession of an eccentric accretion disk on its way to quiescence \citep{zur02}, it is important to determine the orbital inclination angle while XTE J1118+480 is in a truly quiescent state.  

In order to determine an accurate orbital inclination angle for XTE J1118+480, we have obtained $B$-, $V$-, $R$-, $J$-, $H$-, and $K_s$-band light curves of the system while in quiescence, and simultaneously model them here with the WD98 light curve modeling code \citep{wil98}.  To date, this is the most comprehensive ellipsoidal variation data set published for this system.  The modeled inclination angle is combined with recently published orbital parameters to determine a highly constrained mass of the black hole in this X-ray binary.

\section{Observations \& Data Reduction}

We observed the XTE J1118+480 field in the optical and IR
wavelength regimes.  Table~\ref{obstab} summarizes our observations, while we describe them in detail below.

\subsection{Optical Observations}

We obtained optical observations of XTE J1118+480 over three nights in 2003 and four nights in 2004 using
 standard Johnson $B$, $V$, and $R$ filters with the 1.5 m telescope 
at the T\"UB\.ITAK National Observatory (TUG)\footnote{See http://www.tug.tubitak.gov.tr/index.html?en} in Antalya, Turkey.
  The data were obtained with the imaging CCD\footnote{See http://www.tug.tubitak.gov.tr/aletler/SDSU\_imaging\_ccd/} on  2003 June 4-6 
and the ANDOR CCD\footnote{See http://hea.iki.rssi.ru/AZT22/ENG/focal.htm} on 2004 March 18-19 and 2004 April 23-24. A total of 31, 84,
 and 83 images were obtained in the $B$-, $V$- and $R$-bands, respectively. The data were reduced with bias frames, dark current frames, and dome flat fields using standard procedures.
The instrumental magnitudes of XTE J1118+480 were derived through the use of the PSF fitting algorithms, DAOPHOT \citep{Ste87} and ALLSTAR in the MIDAS software package using fifteen PSF stars. In addition, a reference star ($\alpha$=11:18:07.10 and $\delta$=+48:03:53.2) close to our transient source was used to reduce the scintillation effects and derive the calibrated magnitudes for a given filter.  The reference star and target star errors were added in quadrature to produce the final errors on the photometric points. The resulting data were combined into 20 bins per orbit.

\subsection{Infrared Observations}

Infrared data on XTE J1118+480 were obtained on 2003 January 17 and 18
using SQIID (Simultaneous Quad Infrared Imaging Device\footnote{See
http://www.noao.edu/kpno/sqiid/}) on the 2.1 meter telescope at
Kitt Peak National Observatory. We simultaneously obtained measurements 
in the $J$-, $H$-, and $K_s$-bands and kept the
number of counts in each exposure in the linear regime of each chip.  
The data were reduced using tasks in the {\it upsqiid} package in IRAF\footnote{IRAF is distributed by the National Optical Astronomy Observatories, which are operated by the Association of Universities for Research in Astronomy, Inc., under cooperative agreement with the National Science Foundation.}. The images taken at one position were subtracted from the images taken at a slightly offset position to remove the sky, dark current, and any bias level.  We then flat fielded the images using twilight sky flats. 

Aperture photometry was performed on XTE J1118+480 and five nearby field stars.
  Using the IRAF {\it phot} package, a differential light curve for each band was generated. We followed standard error propagation rules for calculating the differential magnitude errors from the target and reference star instrumental magnitudes.  The differential photometric results show that over the course of our 
observations, the comparison stars did not vary more than expected from 
photon statistics. We used photometric images of the \citet{hunt98} AS-11 and AS-18 ARNICA standard star fields, as well as 2MASS field stars to calibrate the data set. As with the optical, the data were combined into 20 bins per orbit.

\section{Results and Discussion}

\begin{figure}
\plotone{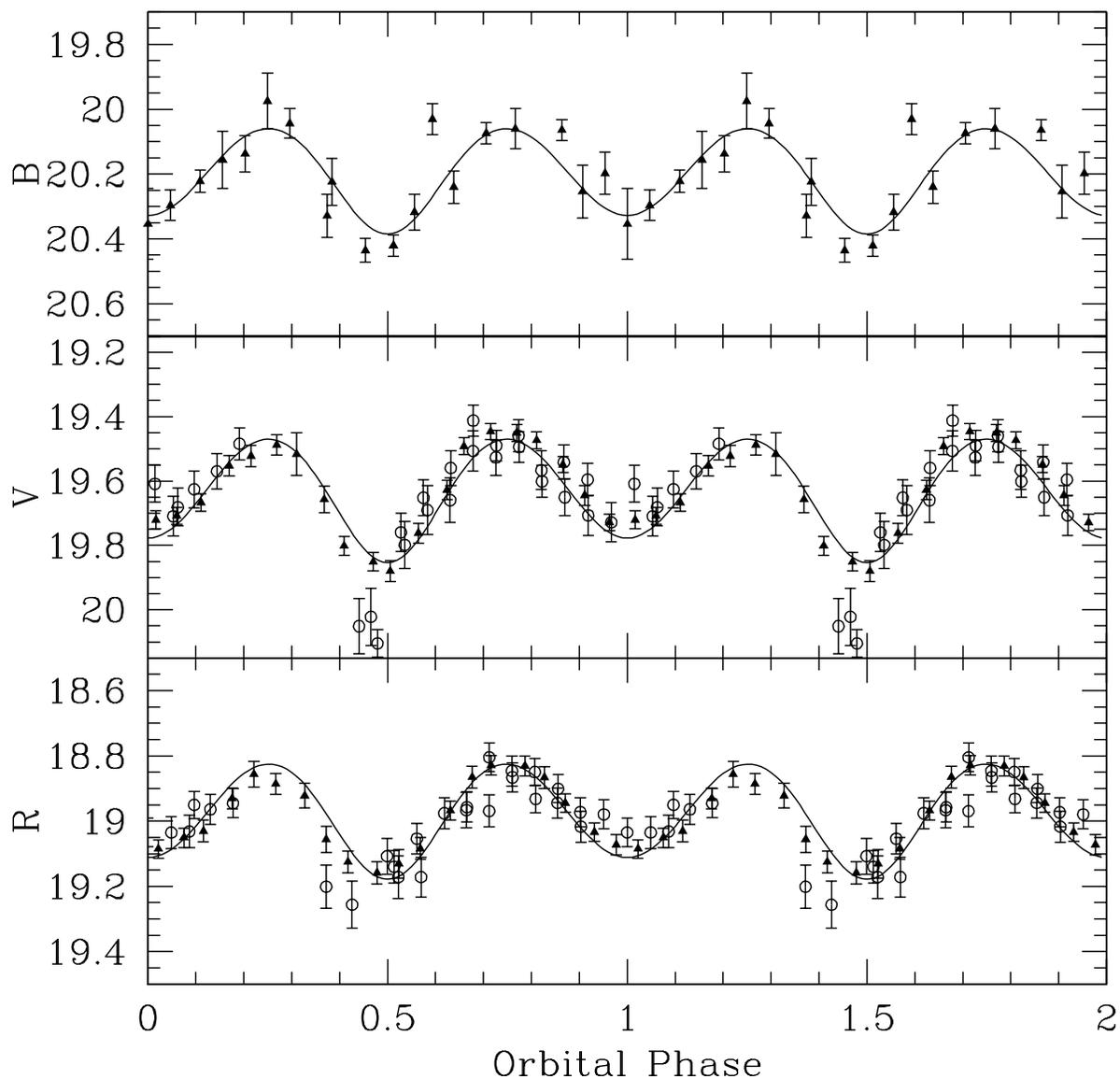}
\caption{XTE~J1118+480 $B$-, $V$-, and $R$-band light curves from 2003 (open circles) and 2004 (filled triangles).  The data are plotted over two phase cycles for clarity. Error bars are 1$\sigma$. The solid line represents the best fitting ($i$ = 68$^{\circ}$) WD98 model as described in the text. \label{fig1}}
\end{figure}

\begin{figure}
\plotone{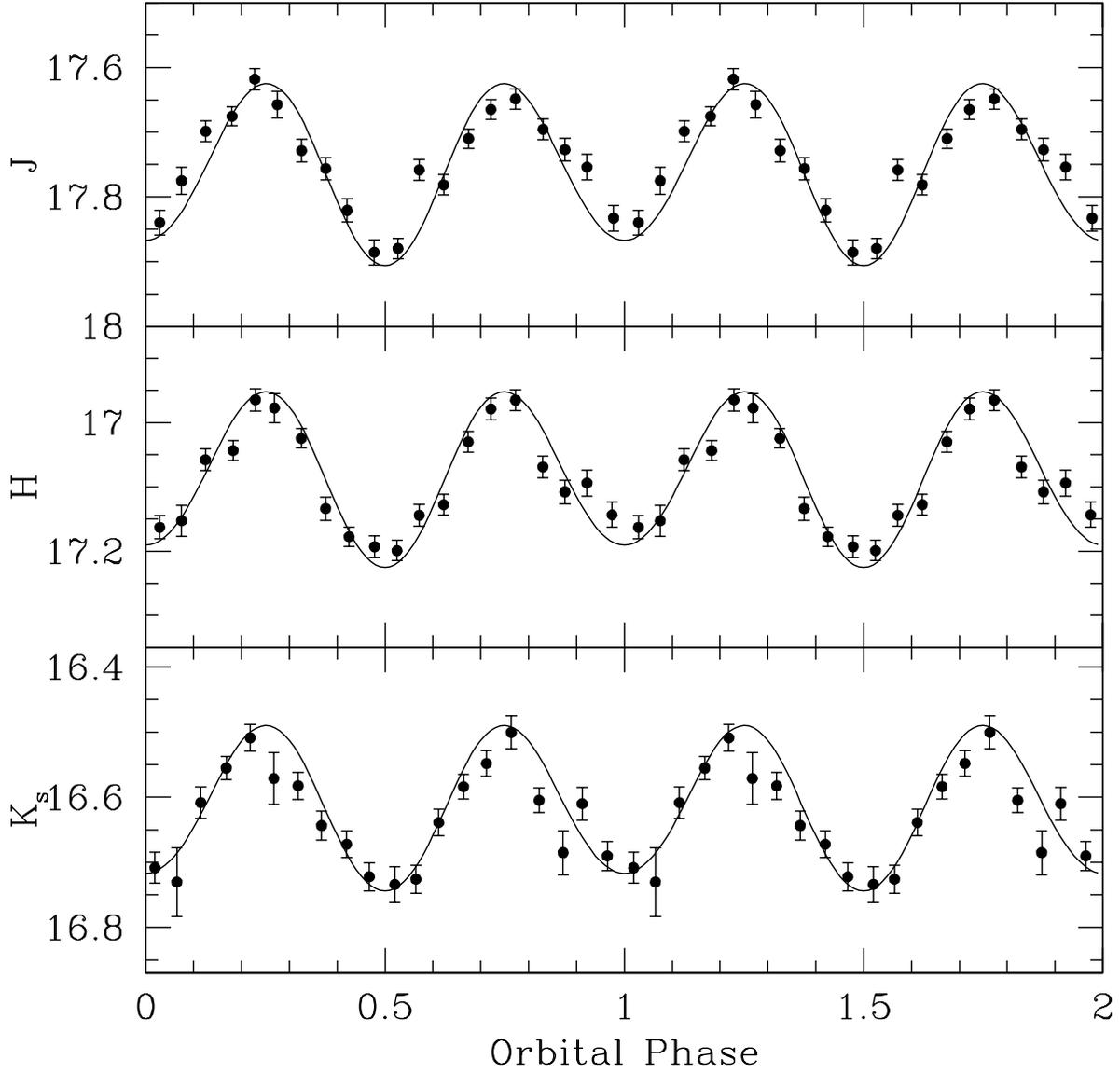}
\caption{XTE~J1118+480 $J$-, $H$-, and $K_s$-band light curves (circles).  The data are plotted over two phase cycles for clarity. Error bars are 1$\sigma$. The solid line represents the best fitting ($i$ = 68$^{\circ}$) WD98 model as described in the text. \label{fig2}}
\end{figure}

Figure \ref{fig1} presents the resulting $B$-, $V$-, and $R$-band light curves of XTE J1118+480, while the final $J$, $H$, and $K_s$ differential light curves are presented in Figure \ref{fig2}. Despite the expectation of detecting superhumps from the 52 day accretion disk precession period, there was no evidence of a superhump period when the data were run through a periodogram.  These results are consistent with the findings of \citet{sha05} whose optical data taken on 2003 June suggested that if a superhump modulation existed, it was at the $<$0.50$\%$ level.  \citet{sha05} also observed stochastic variability and fast flares in their light curves.  However, the observed stochastic variability had the same magnitude as their photometric error bars, and despite the significant power in the power density spectrum, the flares seen in XTE~J1118+480 had roughly the same magnitude as the spread in the comparison star measurements. The 2003 and 2004 $V$- and $R$-band light curves presented here were consistent in shape and amplitude.  As Table \ref{obstab} shows, two thirds of the nights that XTE~J1118+480 was observed, data was gathered for at least 75\% of an orbit, with more than one full orbit covered on four of the nine nights.  No evidence for unequal maxima, flares, or light curve distortions were found in the unphased data.  The periodogram results are consistent with the results from \citet{tor04} and therefore all data presented here have been phased to their ephemeris.  These are the first $H$- and $K_s$-band detections of ellipsoidal variations from this X-ray binary system.

\subsection{Ellipsoidal Models}

\begin{figure}
\plotone{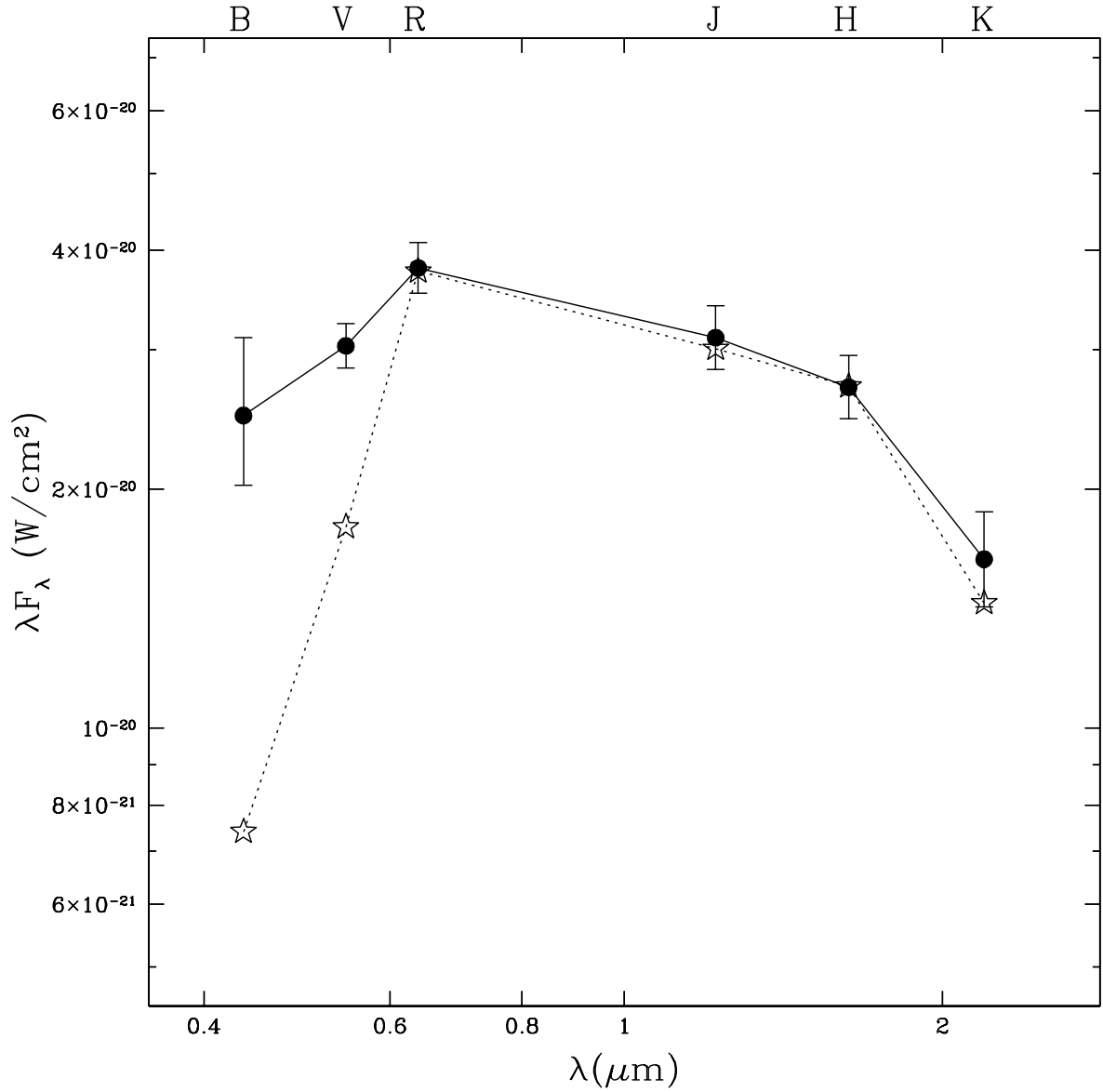}
\caption{XTE~J1118+480 phase-averaged optical-infrared quiescent SED dereddened by $A_V$ = 0.065 mag (filled circles). Error bars are 1$\sigma$. The observed data were compared with SEDs for K0V - M4V stars with A$_V$ = 0.045 - 0.085 mag.  The best fit SED, normalized at $H$, is that of a K7V with 65\% extra light at $B$ and 33\% extra light at $V$ (open stars).  \label{fig3}}
\end{figure}

The spectral type of the secondary star can be estimated by comparing its red optical spectrum with the spectra of stars with various spectral types from the same luminosity class. 
Alternatively, one can use a spectral energy distribution (SED), and published limits on the spectral type to not only derive an effective temperature of
the secondary star, but to also estimate both the visual extinction and 
contamination level. Given that photometric data usually has a higher S/N than spectroscopic data, an effective temperature derived
using photometry can be just as useful as a spectral type derived from a 
spectroscopic data set. To this end, we compared the observed optical/IR ($BVRJHK_s$) SED for 
XTE J1118+480 with the observed SEDs for K0V -- M4V stars \citep{bb88,bes91,mik82}.  We present the best fitting SED in Figure \ref{fig3}, and find that it predicts a visual extinction of A$_V$=0.065$\pm$0.020 mags, and a secondary star spectral type of K7V. It also includes 60 -- 70\% light from the accretion disk at $B$ and 30 -- 35\% at $V$. A K5V gives a slightly worse fit, and predicts more $R$-band light than is observed, as well as a smaller amount of disk light in the $B$- and $V$-bands, inconsistent with previously 
published values. The visual extinction found here is consistent with the column density adopted by \citet[$N_H=1.2 \times 10^{20}$ cm$^{-2}$]{mcc04} based on three independent measurements, and the spectral type found here is consistent with those found through spectral fitting \citep[K5/7 V]{mcc03,tor04}.  

Light from the primary object or accretion disk in an X-ray binary will act to dilute the amplitude of the ellipsoidal variations of the secondary star. \citet{tor04} estimate that the secondary star in the XTE J1118+480 system contributes roughly 55\% of the total flux between 5800\AA ~and 6400\AA; however, at $R$ = 18.6, the system was not in true quiescence when their observations were taken. \citet{zuriauc02} determined that XTE J1118+480 has a true quiescent $R$-band magnitude of 18.9. 
The optical data presented here ($R$=19.00) are consistent with this value, and thus we contend that the system was in a quiescent state during our observations. In addition, Doppler imaging of the system did not detect any H$\alpha$ emission from a hot spot or accretion stream \citep{tor04}. Since our observations took place while XTE J1118+480 was in true quiescence, the optical data presented here are consistent with both of the \citet{tor04} results.

The most difficult contamination source to extract in the case of XTE J1118+480
is the one that has the shallowest spectral slope. If the disk contamination in the infrared is based on the assumption that the optically thin disk radiates through free-free emission processes, and we therefore ascribe the
entire $V$-band excess to free-free emission, then in the $K_s$-band, the
contamination would be $\sim$ 8\%. An 8\% contamination in the infrared bands would cause the observed orbital inclination angle of the system to be underestimated by 2$^{\circ}$. What if we instead assume that any IR contamination is ascribed to a jet or some form of an advection dominated accretion flow (ADAF)? Models by \citet{yua05} for XTE~J1118+480 in quiescence show that both the jet and ADAF flux in the IR are predicted to constitute significantly less than 8\% of the companion flux. 


Irradiation of the secondary star by the accretion disk will affect the symmetry of the ellipsoidal light curves.  \citet{tor04} investigated the possibility of X-ray irradiation powering the H$\alpha$ emission seen in their Doppler tomograms of the XTE J1118+480 system.  They concluded that it was improbable that this could be the case, and found that the strength of the H$\alpha$ emission from the secondary star was comparable with other K dwarfs.  Similarly, we see no significant evidence for irradiation effects in the unphased data or the light curves presented here.

As in \citet{gho01}, we simultaneously modeled the optical and infrared light curves of XTE J1118+480 with WD98 \citep{wil98}.  See \citet{gel01} for references and a basic 
description of the code, and \citet{gelino01} for a more comprehensive 
description. 

The data were run through WD98's DC program which uses a Simplex algorithm for initial parameter searches and a damped least squares algorithm for error minimization between the data and model light curves.  We ran the code for a semi-detached binary with the primary component's gravitational potential set so that all of its mass is concentrated at a point.  
 The most important wavelength-independent input values to WD98 are listed in Table~\ref{tab1}.  The models were run for a range of inclination angles with parameters 
for a K3V through an M1V secondary star.  The secondary star atmosphere was 
determined from solar-metallicity Kurucz models.  We used normal, non-irradiated, square-root limb darkening coefficients \citep{van93}.  We also adopted gravity darkening exponents found by \citet{cla00}, and mass ratios ranging from $q$ = 0.030 - 0.056
\citep{har99,oro01}. We assumed that the secondary 
star was not exhibiting any star spots during our observations. The models were run with varying amounts of additional light in all bands. Solving for six consistent light curve solutions simultaneously allowed the rejection of many disk light scenarios.  

    With 118 degrees of freedom, the best fit model had a reduced $\chi^2$ of 1.65.  While the error on the best fit inclination angle was dependent on combining the uncertainties from varying all of the model parameters simultaneously, we found that changing the spectral type
 of the secondary from a K5V to an M1V resulted in a change in the orbital 
inclination angle of 1$^{\circ}$. Similarly, varying $q$ from 0.030 to 0.056 affected $i$ by $\le 1^{\circ}$.  We find that the best fitting $B$-, $V$-, $R$-, $J$-, $H$-, and 
$K_s$-band model has $i$ = 68$^{\circ}$, and the parameters found in Table \ref{tab2}.  Figure 1 presents this model for the optical bands, while Figure \ref{fig2} presents this model for the $J$-, $H$-, and $K_s$-band light curves.

\subsection{The Model and its Uncertainties}

Based on our optical/IR SED, as well as published values, we adopt a secondary star spectral type of K7 \citep{tor04, mcc01} with a temperature of $T_{\rm eff}$ = 4250 K \citep{gra92}.  The corresponding gravity darkening exponent is $\beta_1$ = 0.34.  

Modeling six light curves simultaneously is a robust method for constraining the amount of extra light in the system. If we model the light curves individually and assume that all of the light in the system comes from the secondary
 star, the best fitting orbital inclination angle for XTE J1118+480 varies with wavelength.  Since the contaminating light does not have the same spectrum as the secondary star the light curves at each wavelength are diluted by different amounts, causing the best fit inclination for the affected bands to be different from the others.  This is the case for the $B$- and $V$-band light curves.  If a jet or other contaminant were to have a flat spectrum that only affected the IR, the results from the SED would most likely not match those obtained from spectral measurements, and we would not be able to determine a reasonable parameter set and inclination angle that is consistent throughout the data. When fit simultaneously, the best fit inclination angle is $i$ = 68$^{\circ}$ with 62\% disk light in $B$, and 31\% disk light in $V$. These disk light contributions are consistent with those found through the SED fitting.

Using an estimate of 8\% for the infrared accretion disk contamination in the system gives an inclination of 
68$^{+2.8}_{-2}$$^{\circ}$, however, if we artificially add in 8\% or more infrared light, we are unable to obtain a reasonable solution for the orbital inclination angle that is consistent throughout the entire data set.  Therefore, based on the IR colors of the system ($J$-$K$=1.1$\pm$0.3), the SED fit, and the results of the simultaneous light curve modeling, it appears unlikely that the infrared 
light curves are significantly affected by any such contamination. In order to determine the final error on the orbital inclination angle, we plotted the $\chi^2$ values as a function of $i$. Therefore, based on the error in each of the model parameters including $q$, the spectral type (i.e. temperature) of the secondary star, the amount of observed disk light, as well as the photometric error bars, the orbital inclination angle is 68$^{\circ}$$\pm$2$^{\circ}$.  
We combined the determined inclination angle with the orbital period (P = 0.1699167 $\pm$ 1.72$\times$10$^{-5}$ d), radial velocity of the secondary star (K$_2$ = 709 $\pm$ 7 km; \citet{tor04}), and and the mass ratio ($q$ = 0.0435 $\pm$ 0.0100) to find the mass 
of the primary object.  A Monte Carlo routine was used to propagate the errors on the above quantities and gives a primary mass of 8.53$\pm$0.60 M$_{\odot}$, confirming it as a black hole. 

The constraints on the mass of the black hole in this system presented here represent a considerable improvement over those previously published. \citet{wag01} determined a mass of 6.0 -- 7.7 M$_{\odot}$ from data obtained before XTE 1118+480 had entered a quiescent state ($R\sim$18.3). \citet{mcc01} gave an upper limit of $M_1 \le$ 10 M$_{\odot}$, and more recently, \citet{mcc04} adopted a mass of $\sim$ 8 M$_{\odot}$ for their calculations of the thermal emission from the black hole in the system.  The 8.53 M$_{\odot}$ black hole mass determined here is consistent with the adopted mass of \citet{mcc04}, and falls nicely into the current observed black hole mass distribution.  Theoretical models by \citet{fry01} predict that there should exist a greater number of 3 -- 5 M$_{\odot}$ black holes than 5 -- 12 M$_{\odot}$ black holes; however, most of the determined black hole masses thus far have fallen into a 6 -- 14 M$_{\odot}$ range.  In fact, thus far, GRO J0422+32 is the only system with a compact object that falls into the 3 -- 5 M$_{\odot}$ range \citep[3.97$\pm$0.95 M$_{\odot}$]{gel03}. 

Using the mass of the compact object and the orbital period, we computed the orbital separation of the two components in the system.  We then combined the separation with the mass ratio to find the size of the Roche lobe for the secondary star.  The temperature of the secondary and its Roche lobe radius were then used to find the secondary's bolometric luminosity and bolometric absolute magnitude. After accounting for the bolometric correction \citep{bes91}, the distance modulus for the $J$, $H$, and $K_s$ bands were used to find an average distance of 1.72$\pm$0.10 kpc. Consistent with results from the modeling and SED fitting, this calculation assumes that all of the IR light in the system originates from the secondary star. Table~\ref{tab2} lists all of the derived parameters for the XTE~J1118+480 system.  Both the mass and radius of the secondary star are smaller than that of a K7V ZAMS star.  In addition, the infrared colors of the system and its position in the Galactic halo, support the notion that the secondary star in the XTE J1118+480 system may be evolved.

\section{Summary}

In this paper, we have presented the first observed $H$- and $K$-band ellipsoidal 
variations, as well as the first $B$-, $V$-, and $R$-band ellipsoidal variations observed while XTE~J1118+480 was in a truly quiescent state.  The derived parameters in Table~\ref{tab2} are 
based on the modeling of these variations including disk contributions of 62\% in the $B$-band and 31\% in the $V$-band. Consistent with \citet{sha05} and \citet{fit03}, we do not see evidence for any optical superhump light or irradiation in the system.  This supports the the fact that the system was in a truly quiescent state during our observations.  

While the orbital inclination angle found here is lower than that found by groups who optically studied the system while approaching quiescence and exhibiting superhumps \citep[$i$=71 -- 82$^{\circ}$]{zur02}, it is consistent with that found from data taken in true quiescence that suggest no significant superhump activity \citep[$i$=63 -- 73$^{\circ}$]{fit03}. Furthermore, the distance we find is consistent with those found previously through both optical \citep[1.9$\pm$0.4 kpc, 1.8$\pm$0.6 kpc, respectively]{wag01, mcc01} and infrared \citep[1.4$\pm$0.2 kpc]{mik05} observations. 

Simultaneously modeling multi-wavelength light curves allows us to better constrain the amount of disk light in an X-ray binary system.  As a result, we have been able to constrain the mass of the black hole in the XTE 1118+480 system to 8.53$\pm$0.60 M$_{\odot}$, and the distance to the system of 1.72$\pm$0.10 kpc.

\acknowledgments
SB acknowledges support from the Turkish Academy of Sciences TUBA-GEBIP (Distinguished Young Scientist) Fellowship. EK acknowledges partial support from T\"UB\.ITAK. UK and SB acknowledge the Turkish National Observatory of T\"UB\.ITAK for providing the facilities for the optical observations. This publication makes use of data products from the Two Micron All Sky Survey, which is a joint project of the University of Massachusetts and the Infrared Processing and Analysis Center/California Institute of Technology, funded by the National Aeronautics and Space Administration and the National Science Foundation.

\clearpage

\begin{deluxetable}{cccc}
\tablecaption{Observations of XTE J1118+480\label{obstab}}
\tablewidth{0pt}
\tablehead{
\colhead{Date} &\colhead{Filter} &\colhead{Exp Time\tablenotemark{a}} &\colhead{\% Orbital Coverage}
}
\startdata
2003 Jan 17 & $J,H,K_s$ & 200 & 179 \\
2003 Jan 18 & $J,H,K_s$ & 200 & 172 \\
2003 Jun 4 & $V,R$ & 240 & 14\tablenotemark{b},9\tablenotemark{c} \\
2003 Jun 5 & $V,R$ & 240 & 62\tablenotemark{b},73\tablenotemark{c} \\
2003 Jun 6 & $V,R$ & 240 & 44\tablenotemark{b},40\tablenotemark{c} \\
2004 Mar 18 & $V$ & 300 & 192 \\
2004 Mar 19 & $R$ & 300 & 224 \\
2004 Apr 23 & $B$ & 420 & 28 \\
2004 Apr 24 & $B$ & 420 & 83 \\
\enddata
\tablenotetext{a} {Effective exposure time per image in seconds}
\tablenotetext{b} {Orbital coverage percentage for the $V$ filter}
\tablenotetext{c} {Orbital coverage percentage for the $R$ filter}
\end{deluxetable}{}
                           
\clearpage

\begin{deluxetable}{lc}
\tablecaption{Wavelength-Independent WD98 Input Parameters\label{tab1}}
\tablewidth{0pt}
\tablehead{
\colhead{Parameter} &\colhead{Value}
}
\startdata
Orbital Period (days) & 0.1699339 \\
Ephemeris (HJD phase 0.0)\tablenotemark{a} & 2451880.1086\\
Orbital Eccentricity & 0.0 \\
Temperature of Secondary (K) & 4250 \\                 
Mass Ratio (M$_2$/M$_1$) & 0.0435 \\
Atmosphere Model & Kurucz (log g = 4.59)\\
Limb Darkening Law & Square-root \\ 
Secondary Star Gravity Darkening Exponent & $\beta_1$=0.34 \\
Secondary Star Bolometric Albedo & 0.676 \\
\enddata
\tablenotetext{a} {From \citet{tor04}}
\end{deluxetable}{}

\clearpage

\begin{deluxetable}{lc}
\tablecaption{Derived Parameters for XTE J1118+480\label{tab2}}
\tablewidth{0pt}
\tablehead{
\colhead{Parameter}&\colhead{Value\tablenotemark{a}}}
\startdata
Amount of Disk Light at $B$ (\%) & 62$\pm$3 \\ 
Amount of Disk Light at $V$ (\%) & 31$\pm$3 \\ 
Orbital Inclination Angle ($^{\circ}$) & 68$\pm$2 \\
Primary Object Mass $M_1$ ($M_{\odot}$) & $8.53\pm 0.60$ \\
Secondary Star Mass $M_2$ ($M_{\odot}$) & $0.37\pm 0.03$ \\
Orbital Separation $a$ ($R_{\odot}$) & $2.67\pm 0.06$ \\
Secondary Star Radius $R_{L_2}$ ($R_{\odot}$) & $0.43\pm 0.01$  \\
Distance (kpc) & $1.72\pm 0.10$ \\
\enddata
\tablenotetext{a} {Errors are 1$\sigma$ ($\Delta \chi^2$ = 1)}
\end{deluxetable}{}

\clearpage

\clearpage

\clearpage

\end{document}